\documentclass[runningheads]{llncs}
\usepackage[T1]{fontenc}
\usepackage{multirow}
\usepackage{caption}
\usepackage{subcaption}
\usepackage{graphicx}
\usepackage{array}
\usepackage{adjustbox}
\usepackage{amsmath,amssymb,amsfonts}
\usepackage{algorithmic}
\usepackage{nccmath}
\usepackage{booktabs,subcaption,amsfonts,dcolumn}
\usepackage{cite}
\usepackage{xcolor}
\usepackage{hyperref}

\hypersetup{
  colorlinks,
  citecolor=blue,
  linkcolor=blue,
  urlcolor=blue}

\def\doi#1{\href{https://doi.org/\detokenize{#1}}{\url{https://doi.org/\detokenize{#1}}}}
\usepackage{graphicx}
\usepackage{listings}
\begin{document}
\title{Deep-ASPECTS: A Segmentation-Assisted Model for Stroke Severity Measurement}

%
\titlerunning{Deep-ASPECTS}
%
\author{Ujjwal Upadhyay\inst{1} \and
Mukul Ranjan\inst{1} \and
Satish Golla\inst{1} \and
Swetha Tanamala\inst{1} \and
Preetham Sreenivas\inst{1} \and 
Sasank Chilamkurthy\inst{1} \and
Jeyaraj Pandian\inst{2} \and
\\Jason Tarpley\inst{3}}
\authorrunning{U. Upadhyay et al.}
%
\institute{Qure.ai\\
\email{\{ujjwal.upadhyay,mukul.ranjan,satish.golla,swetha.tanamala, preetham.sreeniva, sasank.chilamkurthy\}@qure.ai}\\ \and
Christian Medical College Ludhiana, India\\
\email{jeyarajpandian@hotmail.com} \and
Pacific Neuroscience Institute, USA\\
\email{jason.tarpley@providence.org}
}
\maketitle              
\begin{abstract}



A stroke occurs when an artery in the brain ruptures and bleeds or when the blood supply to the brain is cut off. Blood and oxygen cannot reach the brain's tissues due to the rupture or obstruction resulting in tissue death. The Middle cerebral artery (MCA) is the largest cerebral artery and the most commonly damaged vessel in stroke. The quick onset of a focused neurological deficit caused by interruption of blood flow in the territory supplied by the MCA is known as an MCA stroke. Alberta stroke programme early CT score (ASPECTS) is used to estimate the extent of early ischemic changes in patients with MCA stroke. This study proposes a deep learning-based method to score the CT scan for ASPECTS. Our work has three highlights. First, we propose a novel method for medical image segmentation for stroke detection. Second, we show the effectiveness of AI solution for fully-automated ASPECT scoring with reduced diagnosis time for a given non-contrast CT (NCCT) Scan. Our algorithms show a dice similarity coefficient of \textbf{0.64} for the MCA anatomy segmentation and \textbf{0.72} for the infarcts segmentation. Lastly, we show that our model's performance is inline with inter-reader variability between radiologists.

\end{abstract}
\section{Introduction}

A stroke is a medical emergency that requires immediate attention. Brain injury and other consequences can be avoided if intervention is taken early. There are two main types of stroke. \textbf{Hemorrhagic Stroke}. When a blood vessel ruptures, it results in hemorrhagic stroke. Aneurysms or arteriovenous malformations (AVM) are the most common causes of hemorrhagic stroke. \textbf{Ischemic Stroke}.When the blood flow to a portion of the brain is blocked or diminished, brain tissue is deprived of oxygen and nutrients, resulting in an ischemic stroke. Within minutes, brain cells begin to die. Ischemic and hemorrhagic strokes are managed differently since they have different causes and effects on the body. Rapid diagnosis is critical for minimising brain damage and allowing the doctor to treat the stroke with the most appropriate treatment strategy for the type.

\begin{figure}[!ht]
    \centering
    \includegraphics[width=\textwidth]{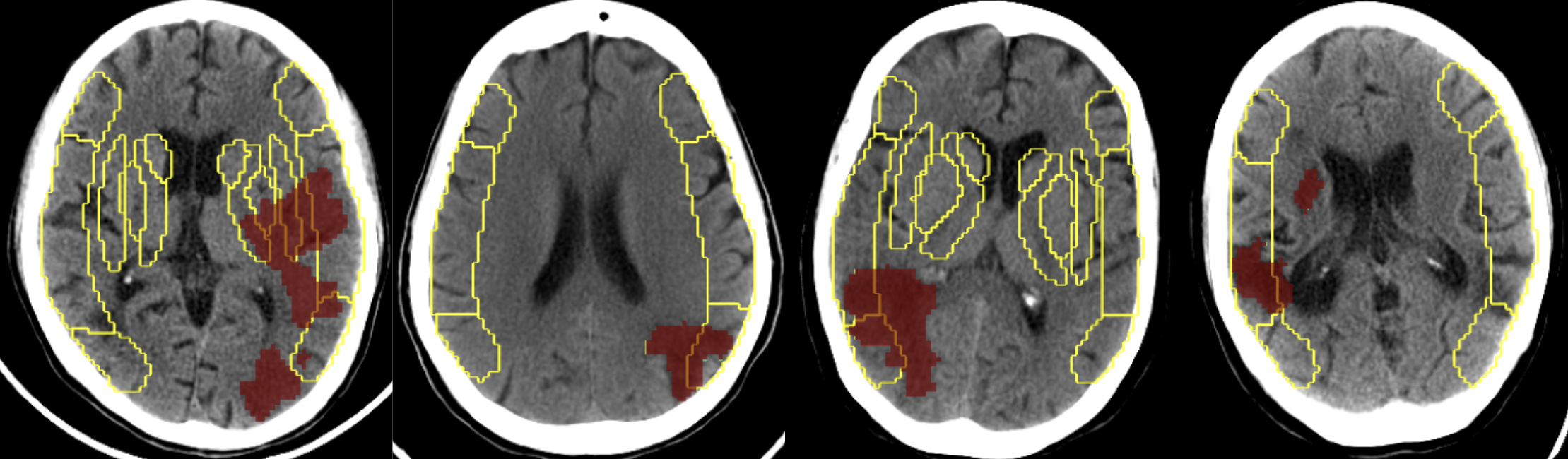}
    \caption{Segmentation output from Deep-ASPECTS. Refer to figure \ref{fig:aspects_pipeline} for more information on how this is generated.}
    \label{fig:aspects_output}
\end{figure}

In this paper, we focus on ischemic strokes. The symptoms of such a stroke vary depending on the brain area affected and the quantity of tissue that has been damaged. The severity of damage can be assessed by two methods "ASPECTS" and "Hypodensity of >1/3 MCA Territory" rule. In clinical practice, ASPECTS detects significant EIC in a higher proportion of the early scans \cite{doi:10.1161/STROKEAHA.117.016745}. ASPECTS is a topographic scoring system that applies a quantitative approach and does not ask physicians to estimate volumes from two-dimensional images. It is scored out of 10 points. ASPECTS has been one of the recognized scoring scales that serve as key selection criteria on the management of acute stroke in the MCA region, where endovascular therapy in patients with baseline ASPECTS$\geq6$ is recommended \cite{doi:10.1161/STR.0000000000000074, esmael_elsherief_eltoukhy_2021, el2017thrombolysis}. Variations of the ASPECT scoring system is used in the posterior circulation and referred to as pc-ASPECTS \cite{doi:10.1161/STROKEAHA.107.511162}. 



In stroke cases, "time is brain"\cite{doi:10.1161/01.STR.0000196957.55928.ab}. The outcomes become progressively worse with time. In the current clinical practice, radiologists must read the NCCT scan to report the ASPECT score, which can take time due to the high volume of cases. We propose an automated ASPECT scoring system, a task currently aﬀected by high inter-reader variability due to manual selection and measurement of the relevant MCA regions scored for ASPECTS. We present Deep-ASPECTS that provides an efficient way to prioritize and detect stroke cases in less than 1 minute, making it clinically relevant.

The proposed deep learning solution involves segmenting the acute infarcts and the MCA territory from an NCCT scan. The segmentation step is followed by overlapping these maps to get the affected region across the slices in basal ganglia and corona radiata level. We demonstrate the robustness of our framework by validating both the segmentation maps and the estimated ASPECTS against ground truth clinical data, which contains 150 scans. 

\section{Related Work}

Various models have been developed and studied for the segmentation of NCCT scan for qualitative estimation of infarcts, haemorrhage and different other critical findings in brain\cite{liang2021symmetry,patel2019intracerebral,kuang2019automated,chilamkurthy2018deep,toikkanen2021resgan}. These studies have presented essential findings related to the NCCT. However, to the best of our knowledge, we could not find any published research which combines the Infarct segmentation and MCA anatomy segmentation to predict the ASPECT score of the NCCT scan in an end-to-end fashion.

Deep learning based semantic segmentation models with encoder and decoder blocks connected through various skip connections have shown to be effective in the medical imaging domain\cite{chen2021transunet, ronneberger2015u,zhou2018unet++,chaurasia2017linknet,hatamizadeh2022unetr}. Unet\cite{ronneberger2015u} presents a simple encoder-decoder based CNN architecture while TransUnet\cite{chen2021transunet} uses transformer\cite{vaswani2017attention,dosovitskiy2020image} based encoder in which a CNN layer is first used to extract the features and then patch embedding is applied to $1\times1$ features map.Linknet\cite{chaurasia2017linknet} is another variation of Unet\cite{ronneberger2015u} which replaces the CNN block of Unet with Residual connection\cite{he2016deep} and instead to stacking the features of encoder blocks to the decoder block in the skip connections it adds them.
UNet++ further extends UNet by connecting encoder and decoder blocks through a series of dense Convolutional blocks instead of directly stacking or adding them.

\begin{figure}[!ht]
    \centering
    \includegraphics[width=\textwidth]{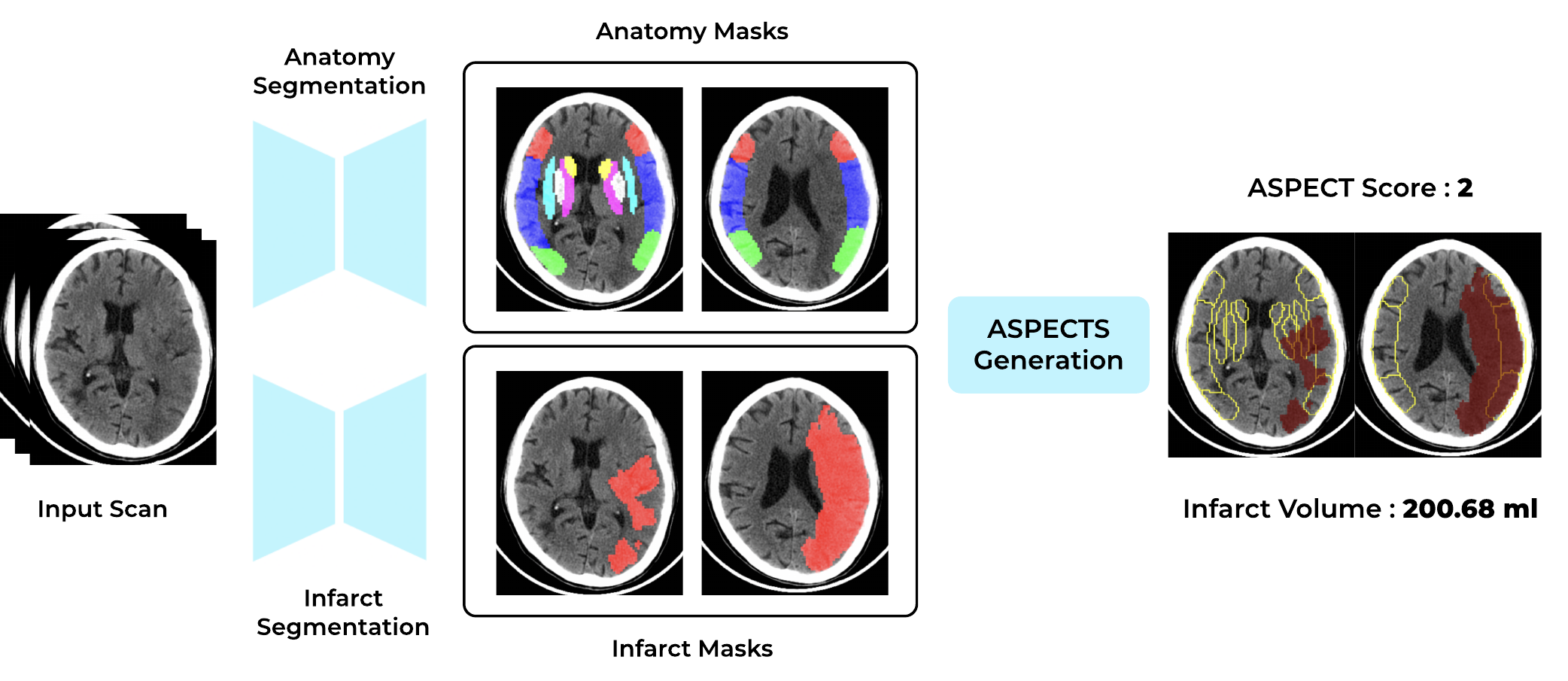}
    \caption{ASPECTS Framework. It has three main components: (1) Infarct segmentation network (2) MCA anatomy segmentation network. (3) ASPECTS generation function.}
    \label{fig:aspects_pipeline}
\end{figure}

\section{Dataset}
We have built our dataset containing 50000 studies out of which 8000 have infarcts. Clinicians manually marked pixel-level ground truths for infarcts and MCA anatomy regions on 1500 training examples. A skilled specialist double-checked the annotation findings under stringent quality control. The labelled dataset was separated into three groups at random in training, validation, and testing at a ratio of 8:1:1. The number of slices in each scan in our dataset varies from subject to subject. Our collaborating hospitals have reviewed the data collection process with approval from the local research ethics committee. Details for data collection process is added to supplementary.


\section{Methods}

The ASPECT score is calculated by dividing the MCA territory into ten regions: Caudate, Lentiform Nucleus, Internal Capsule, Insular Cortex, M1, M2, M3 M4 M5, M6. 

\begin{figure}[!hp]
    \centering
    \includegraphics[scale=0.9]{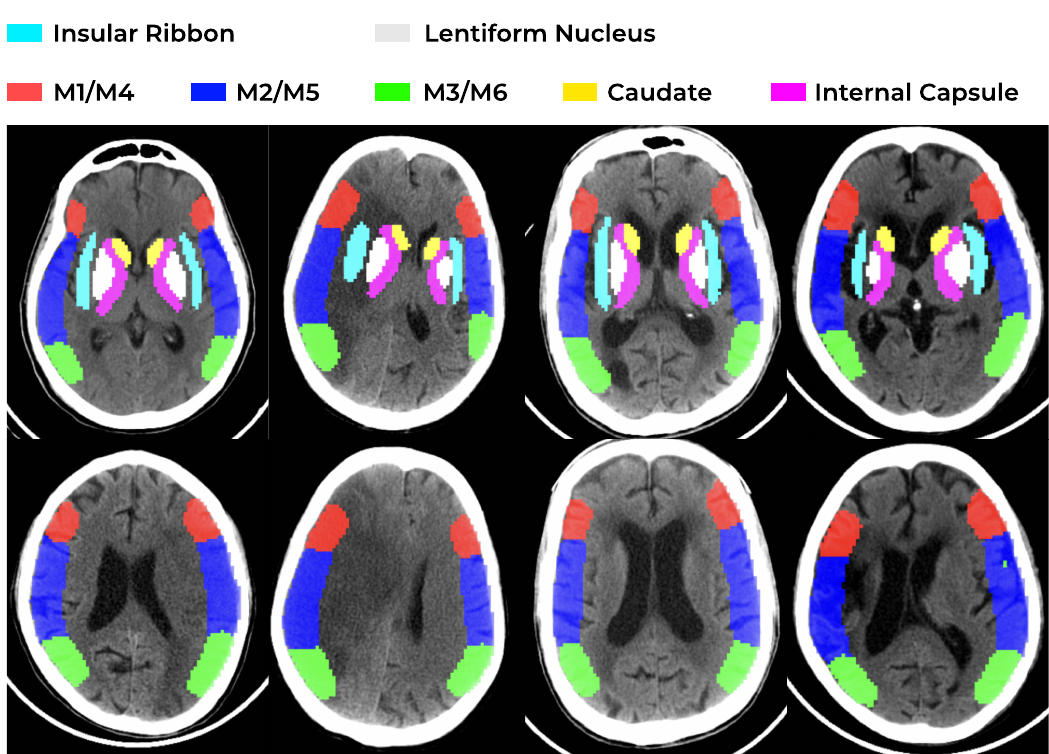}
    \caption{MCA segmentation model results on our test set. These results are based on EfficientNet \cite{DBLP:journals/corr/abs-1905-11946}, achieving a DSC of 0.64. Masks are shown in color.}
    \label{fig:anatomy_masks}
\end{figure}

M1 to M3 is at the level of the basal ganglia. M4 to M6 are at the level of the ventricles immediately above the basal ganglia called corona radiata. 1 point is deducted from the initial score of 10 for every region showing early ischemic signs, such as focal swelling or infarcts. The score was created to aid in identifying patients who were most likely to benefit clinically from intravenous thrombolysis.

Figure \ref{fig:aspects_pipeline} illustrates the pipeline of the proposed framework. The UNet \cite{ronneberger2015u} segmentation models with EfficientNet backbone are used for both infarct and anatomy segmentation networks. ASPECTS generation function is used to overlap segmentation masks from both segmentation networks and report the ASPECT Score. More qualitative results on segmentation mask can be found in figure \ref{fig:aspects_output}. Infarct volume and masked output are also reported to aid the radiologists in making informed decisions regarding following procedures. 

ASPECTS generation function, in figure \ref{fig:aspects_pipeline}, overlays the masks from 2 segmentation over each other and finds the anatomy being overlapped. These overlapped regions are considered affected regions and scored for aspects. This function also finds volume using the `pixel spacing` attribute, part of dicom metadata. Pixel spacing represents how much volume is enclosed by a voxel.

\begin{equation}
\label{eq:volume}
    V(\text{ml}) = \sum{\text{infarct}_{mask}} * \mathit{spacing_x} * \mathit{spacing_y} * \mathit{spacing_z}
\end{equation}

\begin{figure}[!ht]
    \centering
    \includegraphics[scale=0.9]{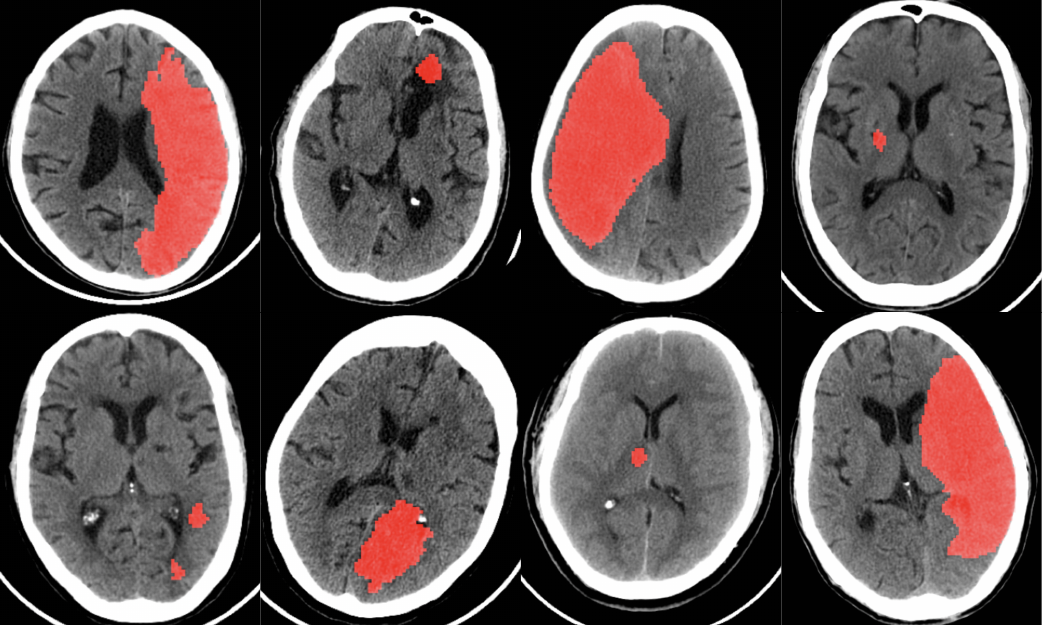}
    \caption{Infarct segmentation model results on our test set. These results are based on EfficientNet \cite{DBLP:journals/corr/abs-1905-11946}, achieving a DSC of 0.72. Infarct masks are shown in red color.}
    \label{fig:infarct_masks}
\end{figure}

Model outputs for infarcts and MCA anatomy are visualized in figures \ref{fig:infarct_masks} and \ref{fig:anatomy_masks}.
Models achieve good results even under challenging conditions when infarct is small and hard to identify. Quantitative results are shown in table \ref{tab:infarct-volume}.

All the scans were rescaled to 224x224 and used as input to our framework. Data augmentation is also applied by introducing random noise, rotation, shift, and flipping. The models were trained on 1 Nvidia 1080Ti GPU with 12GB RAM. We used SGD optimizer with a cyclic learning rate scheduler for all experiments with a learning rate of 5e-3.

\begin{equation}
\label{eq:loss}
    L = \alpha * L_1 + \beta * L_2 + \gamma * L_3
\end{equation}

In equation \ref{eq:loss}, We used 3 loss functions - focal loss \cite{DBLP:journals/corr/abs-1708-02002} ($L_1$), boundary loss \cite{boundaryloss2021} ($L_2$) and dice loss \cite{DBLP:journals/corr/SudreLVOC17} ($L_3$). We found $\alpha=3$, $\beta=1$ and $\gamma=1$, worked best for both infarct segmentation and MCA anatomy segmentation task, after grid search on log linear between 1 and 100. Final evaluation of the acquired data was performed using the dice similarity coeﬃcient (DSC).

\section{Results}

\textbf{How does volume of infarct influence the segmentation map?}. From table \ref{tab:infarct-comparision}, it can be seen that UNet performed better than other models. We also compared the models based on their ability to segment infarct based on their volume. In table \ref{tab:infarct-volume}, it can be seen that UNet was not the best model for infarct volume less than 3ml. UNet++ was better by 0.04 DSC. However, for other volume categories, UNet was still better. Low volume infarct in terms of ASPECTS translates to a higher ASPECT score i.e. more than 8. In table \ref{tab:aspects-with-rads} and \ref{tab:aspects-binned-comparison}, it is evident that the performance is not hampered as even with low dice score. Though in future work, it gives scope of improvement.

\begin{table}[htb]
  \caption{Comparison of different models on infarct segmentation.}
  \scriptsize	
  \begin{subtable}[t]{.45\textwidth}
    \caption{Model Evaluation on infarct segmentation task.}
    \label{tab:infarct-comparision}
    \raggedright
    \begin{tabular}{|c|c|c|c|}
      \hline
      \textbf{Methods} & \textbf{DSC}  & \textbf{Sensitivity} & \textbf{Specificity} \\ \hline
      UNet             & \textbf{0.72} & \textbf{0.77}        & \textbf{0.99}        \\ \hline
      TransUNet        & 0.57          & 0.61                 & 0.98                 \\ \hline
      UNet++           & 0.67          & 0.76                 & 0.99                 \\ \hline
      LinkNet          & 0.54          & 0.57                 & 0.96                 \\ \hline
    \end{tabular}
  \end{subtable}%
  \hspace{0.4cm}
  \begin{subtable}[t]{.5\textwidth}
    \raggedleft
    \caption{Infarct volume used to characterise the performance of the models based on DSC.}
    \label{tab:infarct-volume}
    \begin{tabular}{@{\extracolsep{1pt}}|l|rrrr|}
      \hline
      \multirow{2}{*}{\textbf{Methods}} & \multicolumn{4}{c|}{\textbf{Infarct Volume}}                                             \\ \cline{2-5}
      &
      \multicolumn{1}{l|}{\textbf{\textless{}3ml}} &
      \multicolumn{1}{l|}{\textbf{3-16ml}} &
      \multicolumn{1}{l|}{\textbf{16-66ml}} &
      \multicolumn{1}{l|}{\textbf{\textgreater{}66ml}} \\ \hline
      UNet      & \multicolumn{1}{r|}{0.45}          & \multicolumn{1}{r|}{\textbf{0.72}} & \multicolumn{1}{r|}{\textbf{0.78}} & \textbf{0.91} \\ \hline
      TransUNet & \multicolumn{1}{r|}{0.41}          & \multicolumn{1}{r|}{0.55}          & \multicolumn{1}{r|}{0.61}          & 0.72          \\ \hline
      UNet++    & \multicolumn{1}{r|}{\textbf{0.49}} & \multicolumn{1}{r|}{0.62}          & \multicolumn{1}{r|}{0.71}          & 0.86          \\ \hline
      LinkNet   & \multicolumn{1}{r|}{0.31}          & \multicolumn{1}{r|}{0.54}          & \multicolumn{1}{r|}{0.52}          & 0.79          \\ \hline
    \end{tabular}
  \end{subtable}
\end{table}

\textbf{How do different models segment MCA territory?} Table \ref{tab:mca-models} shows that UNet performed better than other models when we look at the overall DSC of the MCA territory. Though the smaller regions like Lentiforum nucleus, Internal Capsule, and Insular Ribbon UNet is not the best model, we can see that the variation is not very large between UNet and the best models.

\begin{table}[]
\centering
\caption{Comparison of different models on MCA territory segmentation task.The table contains DSC for different regions and across all regions.}
\label{tab:mca-models}
\resizebox{\textwidth}{!}{%
\begin{tabular}{|c|c|c|c|c|c|c|c|c|}
\hline
\textbf{Methods} &
  \textbf{Overall} &
  \textbf{Caudate} &
  \textbf{Lentiform Nucleus} &
  \textbf{Internal Capsule} &
  \textbf{Insular Ribbon} &
  \textbf{M1, M4} &
  \textbf{M2, M5} &
  \textbf{M3, M6} \\ \hline
\textbf{UNet}      & \textbf{0.64} & \textbf{0.70} & 0.60 & 0.59 & 0.52 & 0.72 & \textbf{0.77} & \textbf{0.54} \\ \hline
\textbf{TransUNet} & 0.59 & 0.63 & 0.59 & 0.57 & 0.43 & \textbf{0.73} & 0.71 & 0.49 \\ \hline
\textbf{UNet++}    & 0.62 & 0.69 & \textbf{0.61} & 0.58 & 0.50 & 0.69 & 0.75 & 0.54 \\ \hline
\textbf{LinkNet}   & 0.63 & 0.70 & 0.61 & \textbf{0.61} & \textbf{0.53} & 0.71 & 0.77 & 0.49 \\ \hline
\end{tabular}%
}
\end{table}

\textbf{Model performance for different ASPECT score.} In table \ref{tab:aspects-with-rads}, we can see the performance of the best performing model from infarct segmentation and anatomy segmentation tasks. This performance is measured against two readers. 

From table \ref{tab:aspects-with-rads}, it can be observed that our model has low specificity for some ASPECT scores. However, it is important to notice that specificity is low even between the score of 2 readers. This is attributed to the subjective nature of ASPECTS. Therefore, it is best to judge the performance of model-based on binned ASPECTS, in table \ref{tab:aspects-binned-comparison}. The score was binned based on the treatment outcome given the score. If the score is between 10-8, the treatment outcome is usually favourable and less complex. The ASPECTS cutoff value determined for the prediction of unfavourable outcomes was equal to 7 \cite{esmael_elsherief_eltoukhy_2021}. Patients with ASPECTS < 4 has even less chance of good functional outcome. The improvement in specificity is evident from table \ref{tab:aspects-binned-comparison}. Here we observe that our agreement with Reader B is more than Reader A. The measure of variability between these two readers is discussed in section \ref{sec:inter-reader}.

\begin{table}[]
\centering
\caption{Performance of trained model, Deep-ASPECTS against 2 readers and inter-reader difference for each ASPECTS.}
\label{tab:aspects-with-rads}
\resizebox{\textwidth}{!}{%
\begin{tabular}{|c|cc|cc|cc|}
\hline
\multirow{2}{*}{\textbf{ASPECTS}} &
  \multicolumn{2}{c|}{\textbf{Model vs Reader A}} &
  \multicolumn{2}{c|}{\textbf{Model vs Reader B}} &
  \multicolumn{2}{c|}{\textbf{Reader A vs Reader B}} \\ \cline{2-7} 
 &
  \multicolumn{1}{c|}{\textbf{Sensitivity}} &
  \textbf{Specificity} &
  \multicolumn{1}{c|}{\textbf{Sensitivity}} &
  \textbf{Specificity} &
  \multicolumn{1}{c|}{\textbf{Sensitivity}} &
  \textbf{Specificity} \\ \hline
0  & \multicolumn{1}{c|}{0.99} & 0.17 & \multicolumn{1}{c|}{0.98} & 0.67 & \multicolumn{1}{c|}{0.96} & 0.50 \\ \hline
1  & \multicolumn{1}{c|}{1.00} & 0.50 & \multicolumn{1}{c|}{1.00} & 0.50 & \multicolumn{1}{c|}{0.99} & 0.50 \\ \hline
2  & \multicolumn{1}{c|}{1.00} & 0.20 & \multicolumn{1}{c|}{0.99} & 0.20 & \multicolumn{1}{c|}{0.99} & 1.00 \\ \hline
3  & \multicolumn{1}{c|}{0.94} & 0.38 & \multicolumn{1}{c|}{0.99} & 0.50 & \multicolumn{1}{c|}{0.99} & 0.27 \\ \hline
4  & \multicolumn{1}{c|}{0.99} & 0.25 & \multicolumn{1}{c|}{0.98} & 0.25 & \multicolumn{1}{c|}{0.97} & 0.00 \\ \hline
5  & \multicolumn{1}{c|}{0.97} & 0.33 & \multicolumn{1}{c|}{0.96} & 0.83 & \multicolumn{1}{c|}{0.94} & 0.33 \\ \hline
6  & \multicolumn{1}{c|}{0.95} & 0.22 & \multicolumn{1}{c|}{0.97} & 0.22 & \multicolumn{1}{c|}{0.96} & 0.27 \\ \hline
7  & \multicolumn{1}{c|}{0.92} & 0.40 & \multicolumn{1}{c|}{0.89} & 0.30 & \multicolumn{1}{c|}{0.92} & 0.47 \\ \hline
8  & \multicolumn{1}{c|}{0.90} & 0.20 & \multicolumn{1}{c|}{0.92} & 0.30 & \multicolumn{1}{c|}{0.92} & 0.25 \\ \hline
9  & \multicolumn{1}{c|}{0.81} & 0.30 & \multicolumn{1}{c|}{0.69} & 0.40 & \multicolumn{1}{c|}{0.73} & 0.52 \\ \hline
10 & \multicolumn{1}{c|}{0.85} & 0.59 & \multicolumn{1}{c|}{0.93} & 0.35 & \multicolumn{1}{c|}{0.92} & 0.41 \\ \hline
\end{tabular}%
}
\end{table}

\begin{table}[]
\centering
\caption{Performance comparison across binned ASPECT score.}
\label{tab:aspects-binned-comparison}
\resizebox{\textwidth}{!}{%
\begin{tabular}{|c|cc|cc|cc|}
\hline
\multirow{2}{*}{\textbf{ASPECTS}} &
  \multicolumn{2}{c|}{\textbf{Model vs Reader A}} &
  \multicolumn{2}{c|}{\textbf{Model vs Reader B}} &
  \multicolumn{2}{c|}{\textbf{Reader A vs Reader B}} \\ \cline{2-7} 
 &
  \multicolumn{1}{c|}{\textbf{Sensitivity}} &
  \textbf{Specificity} &
  \multicolumn{1}{c|}{\textbf{Sensitivity}} &
  \textbf{Specificity} &
  \multicolumn{1}{c|}{\textbf{Sensitivity}} &
  \textbf{Specificity} \\ \hline
A (0-3)  & \multicolumn{1}{c|}{0.96} & 0.48 & \multicolumn{1}{c|}{0.99} & 0.65 & \multicolumn{1}{c|}{0.95} & 0.56 \\ \hline
B (4-7)  & \multicolumn{1}{c|}{0.87} & 0.55 & \multicolumn{1}{c|}{0.85} & 0.66 & \multicolumn{1}{c|}{0.84} & 0.66 \\ \hline
C (8-10) & \multicolumn{1}{c|}{0.74} & 0.93 & \multicolumn{1}{c|}{0.74} & 0.86 & \multicolumn{1}{c|}{0.84} & 0.85 \\ \hline
\end{tabular}%
}
\end{table}

\textbf{Number of parameters for models and their computation time.}
Table \ref{tab:parameters} shows that other that TransUNet, all of the other three models  UNet, UNet++ and LinkNet have a comparable number of parameters since they have the same backbone of EfficientNet\cite{DBLP:journals/corr/abs-1905-11946}. Despite the fact that UNet has 0.9M more parameters compared to LinkNet, its performance on infarct and MCA anatomy segmentation was better than LinkNet. Computational time was also comparable for both models. Therefore we found UNet to be a better option than LinkNet. This evaluation is done on 1 Nvidia 1080Ti GPU with 12GB RAM.

\begin{table}[]
\centering
\caption{Total number of parameters for each model and time taken to process each scan.}
\label{tab:parameters}
\begin{tabular}{|c|c|c|}
\hline
\textbf{Method} & \textbf{Parameters} & \textbf{Time (seconds)} \\ \hline
UNet            & 42.1M               & 28-30                   \\ \hline
TransUNet       & 105M                & 50-63                   \\ \hline
UNet++          & 42.9M               & 28-32                   \\ \hline
LinkNet         & 41.2M               & 27-29                   \\ \hline
\end{tabular}
\end{table}

\section{Inter-Reader Variability}
\label{sec:inter-reader}
Out of 1500 scans, 150 scans were read by two radiologists. This set is the same as the test set to learn what the acceptable ASPECT score might be for a scan. In Tables \ref{tab:aspects-with-rads} and \ref{tab:aspects-binned-comparison}, we compare how much disagreement there is between 2 readers with regards to ASPECTS from 10 and a binned version of score. 

ASPECTS is a very subjective score, and it varies from reader to reader. We found the inter-reader agreement on our data to be 39.45\% when we expect an exact score match. However, the agreement increases to 76.87\% when the difference of 2 points is allowed. Pearson correlation between the reads of 2 readers came out to be 73.17\%.

The ASPECTS reported by our model have an agreement of 42.17\% and 36.73\% with readers A and B, respectively. If the agreement score is relaxed by 2 points, then the agreement increases to 69.38\% and 76.19\%. These numbers suggest that our deep learning model performed closely with radiologists maintaining comparable agreement scores.

\section{Conclusion}
We proposed a novel end-to-end system for automated ASPECT scoring. Additionally, a pilot study was conducted, showing the effectiveness of such an AI-based model in getting ASPECTS promptly. Deploying a system that can score an NCCT in less than a minute with reasonable accuracy can aid preliminary diagnosis for a suspected stroke case. Deep-ASPECTS can save precious time from the diagnosis phase and prevent further deterioration in patients' conditions. 

Deep-ASPECTS got an agreement of 76.19\% with reader B (radiologists), which is the same as reader A (refer section \ref{sec:inter-reader}). The proposed method consistently performs as good as radiologists. In sum, the method shows a high potential to improve the clinical success rate by alerting the radiologist or neurologists about potential stroke cases with their severity reported as ASPECTS. Furthermore, a more realistic study should be conducted with a bigger sample size. In future, we intend to do a more thorough architecture search for the encoders, investigate more anatomical priors, and improve the model performance even further.


\bibliography{main}
\end{document}